\documentclass[twocolumn, superscriptaddress, preprintnumbers, amsmath, amssymb, aps, physrev]{revtex4-2}
\usepackage{graphicx}
\usepackage{dcolumn}
\usepackage{bm}
\usepackage{hyperref}
\usepackage{nameref}
\usepackage{cleveref}
\usepackage{physics}
\usepackage{subcaption}
\usepackage{bbold}
\usepackage{caption}
\usepackage{dsfont}
\usepackage{color}
\usepackage{pdfcomment}
\begin{document}
	\setcitestyle{sort=false}
	\title{\textbf{Effect of Two-Body Interactions on Floquet topological phases}}
	\author{Arijit Dutta}
	\email{arijitd14@gmail.com}
	\affiliation{Goethe-Universität, Institut für Theoretische Physik, 60438 Frankfurt am Main, Germany}
	\author{Souradeep Roy Choudhury}
	\email{roychoudhury@physik.uni-frankfurt.de}
	\affiliation{Goethe-Universität, Institut für Theoretische Physik, 60438 Frankfurt am Main, Germany}
	\author{Tao Qin}
	\affiliation{School of Physics, Anhui University, Hefei, Anhui Province 230601, People’s Republic of China}
	\author{Walter Hofstetter}
	\email{hofstett@physik.uni-frankfurt.de}
	\affiliation{Goethe-Universität, Institut für Theoretische Physik, 60438 Frankfurt am Main, Germany}
	\date{\today}
	\begin{abstract}\noindent
		We study the circularly driven Falicov-Kimball model on a honeycomb lattice within real space Floquet dynamical mean field theory (DMFT). The noninteracting version of this model has been realized experimentally~\cite{wintersperger2020realization}. The noninteracting system hosts an effective Haldane phase at large driving frequencies, while at intermediate frequencies it hosts an anomalous topological phase. We study the effect of two-body interactions $U$ on the stability of these phases. We find that charge pumping does not remain quantized upon increasing $U$, despite the presence of edge modes in the spectrum. This can be attributed to the broadening of the edge modes due to interaction. We also calculate the rate of energy dissipation into the bath and find remarkably different behaviour in the two regimes.
			\begin{description}
				\item[Keywords]
				Floquet Systems, Floquet Topological Phases, Anomalous Floquet Topological Insulator, Dynamical Mean-Field Theory, Strongly Correlated Systems.
			\end{description}
	\end{abstract}
	\maketitle
	\section{Introduction}\noindent
		\noindent Quantum phases of matter with nontrivial topology and inherent robustness to perturbations are a major area of current research. Time-periodically driven ultracold atoms in optical lattices have emerged as a versatile and powerful platform for realizing and studying these phases \cite{goldman2014periodically, goldman2016topological, gross2017quantum, langen2015ultracold, cooper2019topological, hofstetter2018quantum}. Two paradigmatic models, the Hofstadter model and the Haldane model have been realized using Raman laser assisted tunneling \cite{aidelsburger2013realization} and hexagonal lattice shaking \cite{jotzu2014experimental} respectively. Different techniques have been developed and applied in order to measure band topology experimentally. Measurement of the transverse drift in response to a constant applied force enabled the experimental quantification of the Chern number corresponding to the lowest band \cite{aidelsburger2015measuring}. The Chern number of the lowest band was also determined through a transverse drift measurement upon subjecting the lattice to a movement along a periodic trajectory, giving rise to an inertial restoring force exerted on the atoms \cite{jotzu2014experimental}. \par\noindent
		Time-periodic driving introduces novel topological properties due to a periodic quasienergy spectrum \cite{eckardt2015high}. In the limit of high frequencies, the system can be modeled by an effective time-independent Hamiltonian \cite{eckardt2015high}. However, for intermediate driving frequencies, comparable to other energy scales of the system, anomalous topological phases emerge which cannot be described using single particle Floquet band Chern numbers \cite{kitagawa2010topological, liu2010quantum, rudner2013anomalous, wintersperger2020realization, dutta2024anomalous, zheng2024floquet}. For instance, the anomalous Floquet topological insulator (AFTI) supports chiral edge modes within all bandgaps despite a topologically trivial bulk \cite{kitagawa2010topological, rudner2013anomalous, zheng2024floquet}. \par\noindent
		Introducing two-particle interactions into a time-periodically driven system makes the problem highly non-trivial. Experimentally, heating becomes a major problem. Such a system is expected to heat up to a trivial state with infinite temperature \cite{d2014long, lazarides2014equilibrium, mallayya2019heating}, with the exception of many-body localized systems \cite{mondaini2015many}, integrable systems \cite{bertini2015prethermalization} and the prethermalization plateau \cite{bukov2015prethermal}. Multi-photon interband heating has been observed in a shaken 1D optical lattice \cite{weinberg2015multiphoton}. Floquet evaporative cooling was shown to reduce heating for interacting bosons in a one-dimensional optical lattice \cite{reitter2017interaction}. However, there are no artificial gauge fields in these setups. Further efforts are needed to go into the interacting regime and realize an interacting system with artificial gauge fields. Theoretically, perturbative methods and effective static Hamiltonian-based approaches are often insufficient to capture the effects of strong correlations or of intermediate driving frequencies~\cite{eckardt2015high}. \par\noindent
		Dynamical mean field theory (DMFT) is inherently nonperturbative with respect to the interaction strength and can capture the effects of strong correlations, such as the Mott insulator transition for intermediate interaction strengths, exactly in the limit of infinite spatial co-ordination number \cite{georges1992hubbard}. In order to study the nonequilibrium steady states (NESS) of periodically driven strongly correlated systems, DMFT is combined with the Floquet-Keldysh formalism \cite{keldysh1964diagram, schmidt2002nonequilibrium, freericks2006steady, shvaika2006resonant, tsuji2008correlated, aoki2014nonequilibrium, 2011theoretical}. For a system in the presence of extrinsic dissipation, a NESS independent of initial correlations both within the system as well as between the system and the bath has been shown to exist \cite{tsuji2009nonequilibrium}. Thus, we couple our system to an explicit heat bath. \par\noindent
		We use the real-space version of Floquet DMFT (RFDMFT) \cite{qin2017spectral, qin2018charge}, which enables us to investigate driven inhomogeneous systems with nontrivial topology in the presence of interactions. We study the Falicov-Kimball model for ultracold fermions on a time-periodically modulated two-dimensional hexagonal lattice coupled to a noninteracting fermionic bath in equilibrium at a finite temperature. For studying the bulk spectrum, we impose periodic boundary conditions along both spatial dimensions. For evaluating the topological properties of the phases, we consider a cylinder geometry in the presence of a flux threaded through the cylinder axis. Varying the threaded flux over the range of a single flux quantum in a topologically nontrivial phase leads to a discontinuity in the pumping of fermions from one edge of the cylinder to the other \cite{laughlin1981quantized}. The pumped charge becoming very small signals a topological phase transition to the topologically trivial Mott insulator phase at large interaction strengths \cite{qin2018charge}. We also calculate the local density of states (LDOS) for commenting on the nature of such a phase transition from a topologically nontrivial phase to the topologically trivial Mott insulating phase in more detail. Since dissipation also plays a role in the stability of the Floquet topological phases, we study dissipation explicitly employing an exactly solvable heat bath. \par\noindent
		The manuscript is organized as follows: In Section \ref{section_model}, we describe the model employed. In Section \ref{section_rfdmft}, we explain the RFDMFT method for the driven interacting system. 
		In Section \ref{section_bulk_spectrum} and \ref{section_edge_spectrum}, we present our results on the bulk and the edge spectrum of the system respectively. In Section \ref{section_charge_pump}, we investigate the charge pumped in a Laughlin pump setup. In Section \ref{section_ldos}, we comment on the LDOS. In order to investigate the potential heating of the system, we discuss the rate of dissipation of energy from the system to the bath in Section \ref{section_open_system}. We conclude in Section \ref{section_conclusion}.
	\section{Model}
	\label{section_model}
\begin{figure}[!th]
			\begin{subfigure}{0.35\linewidth}
				\centering
				\includegraphics[width = \linewidth, keepaspectratio]{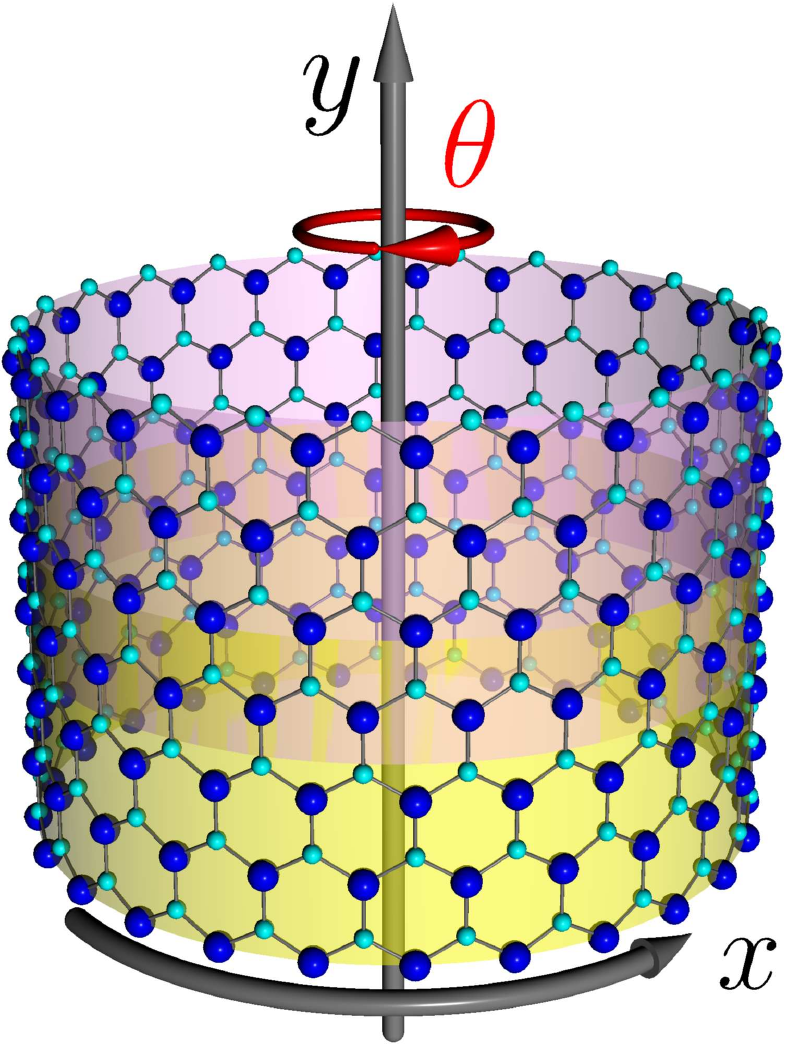}	
				\caption{}		
				\label{figure_cylinder_lattice}
			\end{subfigure}
			\hfill
			\begin{subfigure}{0.6\linewidth}
				\centering
				\includegraphics[width = \linewidth, keepaspectratio]{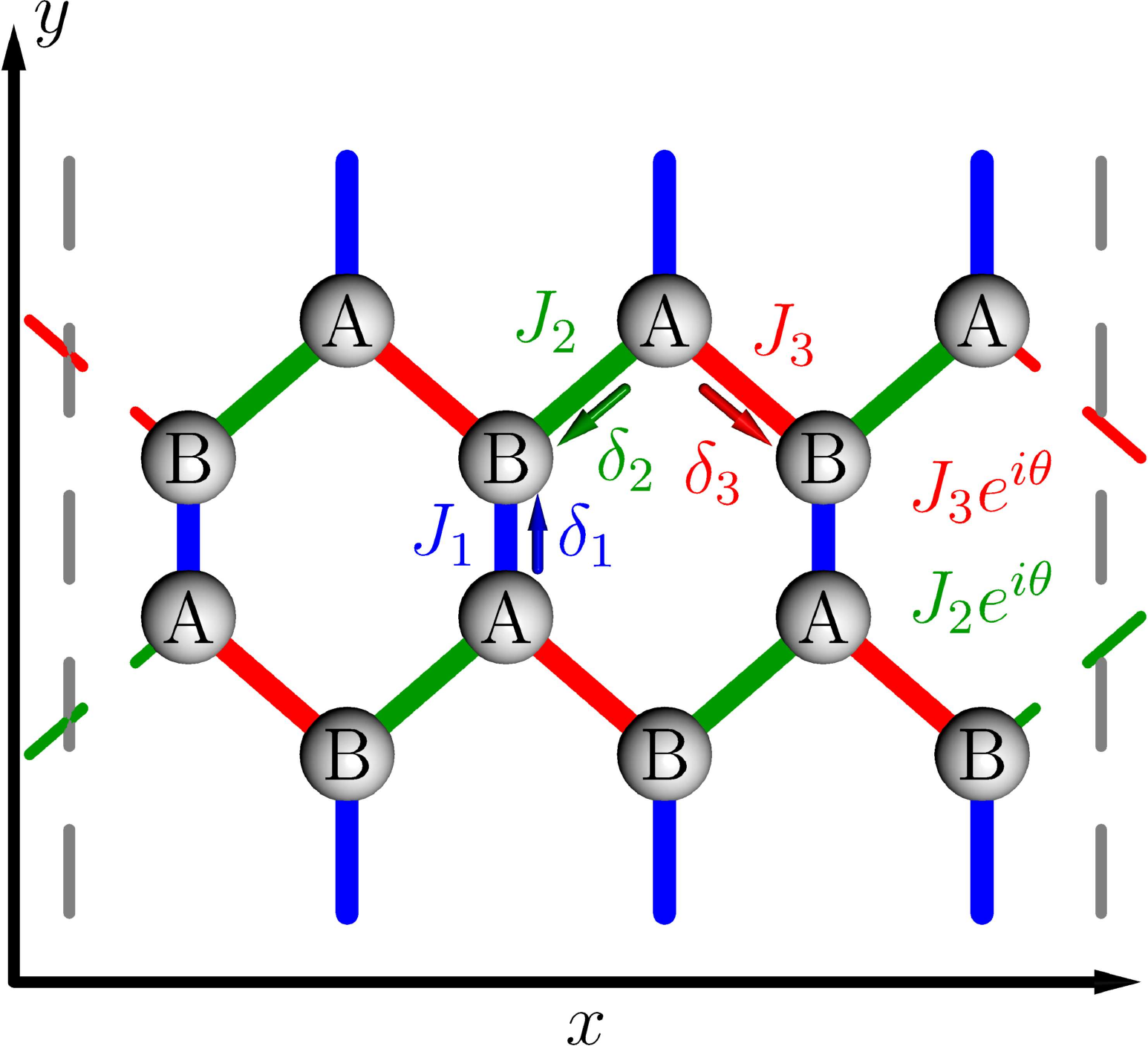}	
				\caption{}
				\label{figure_lattice_2D}
			\end{subfigure}
			\caption{\subref{figure_cylinder_lattice}) Cylinder geometry in the presence of a threaded flux $\theta$. The lattice-sites are denoted by black spheres connected by black lines. The lower (upper) half is represented by yellow (pink) solid fill. \subref{figure_lattice_2D}) The honeycomb lattice. We denote lattice-sites by gray filled spheres, along with the sublattice A or B written on the sphere. We represent the hopping strengths $J_{1}\left(t\right)$ along $\pmb{\delta}_{1}$ with blue, $J_{2}\left(t\right)$ along $\pmb{\delta}_{2}$ with red and $J_{3}\left(t\right)$ along $\pmb{\delta}_{3}$ with green lines respectively. The gray dashed lines at the left and right edges represent the spatial boundaries along the line (1, 0) in two-dimensional Cartesian co-ordinates, while the green and red dashed lines crossing these boundaries represent the choice of gauge for which $J_{2}$ and $J_{3}$ acquire a phase factor $\exp\left(i\theta\right)$ respectively, where $\theta$ is the threaded flux.}
		\end{figure}

\noindent
		\noindent We consider a two-species fermionic mixture on a two-dimensional time-periodically driven hexagonal lattice at half-filling, i.e., with an average of one fermion per lattice-site. The fermions experience time-periodic nearest-neighbour tunneling and onsite two-particle interactions described by the Falicov-Kimball model \cite{falicov1969simple}. Of the two species, the itinerant fermions, created (annihilated) at lattice-site $\pmb{i}$ by $\hat{c}^{\dagger}_{\pmb{i}}\!\left(\hat{c}^{\vphantom{\dagger}}_{\pmb{i}}\right)$, can tunnel throughout the lattice. However, the immobile fermions, created (annihilated) at lattice-site $\pmb{i}$ by $\hat{f}^{\dagger}_{\pmb{i}}\!\left(\hat{f}^{\vphantom{\dagger}}_{\pmb{i}}\right)$, do not tunnel. Species-selective tunneling of this type has been realized experimentally using ultracold atoms in an optical lattice by applying a magnet field gradient modulated in time \cite{jotzu2015creating}. We use the single-band approximation, valid for unpolarized spin-half fermions for all fillings in which there exist on average, not more than two fermions per lattice-site, under low temperatures for sufficiently deep optical lattices, and neglect longer range tunneling than nearest neighbour tunneling \cite{hofstetter2018quantum}. The Hamiltonian describing the lattice system at time $t$ is given by
		\begin{equation}
			\begin{split}
				\mathcal{H}_{S}\!\left(t\right) = & \mathcal{H}_{0}\!\left(t\right) + \mathcal{H}_{int}\\
				\mathcal{H}_{0}\!\left(t\right) = & -\sum\limits_{\gamma = 1}^{3}\sum\limits_{\bm{i}}J_{\gamma}\!\left(t\right)\left(\hat{c}^{\dagger}_{\pmb{i}}\hat{c}^{\vphantom{\dagger}}_{\pmb{i} + \pmb{\beta}_{\pmb{i}\gamma}} + h. c.\right)\\ 
				\mathcal{H}_{int} = & U\sum\limits_{\pmb{i}}\hat{c}_{\pmb{i}}^{\dagger}\hat{c}_{\pmb{i}}^{\vphantom{\dagger}}\hat{f}_{\pmb{i}}^{\dagger}\hat{f}_{\pmb{i}}^{\vphantom{\dagger}},
			\end{split}
		\end{equation}\noindent
		where $J_{\gamma}\!\left(t\right)$ denotes the nearest-neighbour hopping strengths. The lattice is bipartite and we denote the sublattices by A and B. If the lattice-site $\pmb{i}$ is in the A (B) sublattice, $\pmb{\beta}_{\pmb{i}\gamma} = + (-)\pmb{\delta}_{\gamma}$ for $\gamma \in \left\{1, 2, 3\right\}$. In two-dimensional Cartesian co-ordinates, $\pmb{\delta}_{1} \equiv \left(0, a\right)$, $\pmb{\delta}_{2} \equiv \left(-\sqrt{3}a/2, -a/2\right)$ and $\pmb{\delta}_{3} \equiv \left(\sqrt{3}a/2, -a/2\right)$, where $a$ is the lattice constant (see Fig. \ref{figure_lattice_2D}). We consider $a = 1$. \par\noindent
		\par
\begin{figure}[!h]
			\centering
			\includegraphics[width = \linewidth, keepaspectratio]{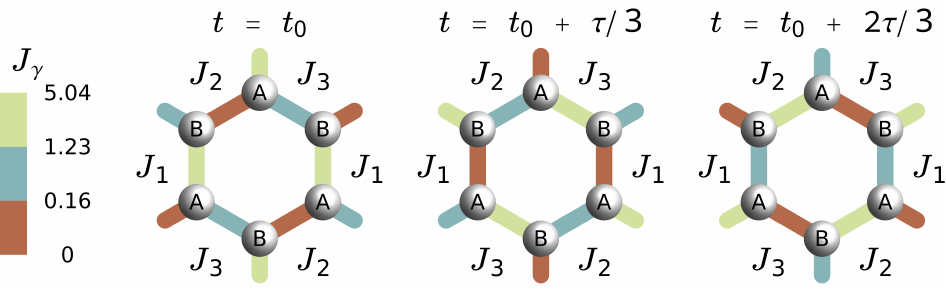}	
			\caption{The driving protocol for the hopping strengths. We denote lattice-sites by gray filled circles whereas we represent the hopping strengths at different time instants in a false color plot.}
			\label{figure_driving_protocol}
		\end{figure}

\noindent
		We drive the hopping strengths $J_{\gamma}\left(t\right)$ according to the following protocol, which in \cite{dutta2024anomalous} has been found to give rise to the anomalous Floquet Anderson insulator (AFAI) for noninteracting fermions in the presence of quenched disorder. 
		\begin{equation}
			\begin{split}
				J_{\gamma}\!\left(t\right) = & J\exp\left[B\cos\!\left(\Omega t + \phi_{\gamma}\right)\right]\\	
				\phi_{\gamma} = & \dfrac{2\pi\!\left(\gamma - 1\right)}{3}.
			\end{split}
		\end{equation}\noindent
		Here $J$ is the amplitude of the time-periodic modulation of the hopping strength. $B$ is a dimensionless parameter which controls the width of the bulk bands and also determines the driving frequency for transition from the anomalous phase to the Haldane phase in the clean system in the absence of disorder \cite{dutta2024anomalous}. $\Omega$ is the driving frequency. In our numerical simulations presented in the following, we choose the values $J = 1$ and $B / J = 2$. Under this scheme, the three hopping strengths peak alternately, in intervals of $\tau / 3$ in one driving period $\tau$, where $\Omega\tau = 2\pi$, as illustrated in Fig. \ref{figure_driving_protocol}.
		\par\noindent
		An onsite two-particle interaction $U$ couples the itinerant fermions to the immobile fermions at all lattice-sites \cite{falicov1969simple}. This model is equivalent to a model with annealed disorder in equilibrium \cite{eckstein2009nonequilibrium}. At half filling, the two-particle interaction has been shown to lead to a Mott transition \cite{brandt1989thermodynamics, falicov1969simple, freericks2003exact, qin2018nonequilibrium, nguyen2013correlation}. Since the immobile fermions do not tunnel across the lattice, the impurity solver for this model becomes exact, in the limit of infinite spatial co-ordination number \cite{brandt1989thermodynamics} (for details, please consult section \ref{section_rfdmft}). This remains the case even when the system is driven away from equilibrium \cite{eckstein2009nonequilibrium}. Thus, it can be used as a benchmark for nonequilibrium methods as well as a model to study ultracold fermions.\par\noindent
		To study the bulk, we impose periodic boundary conditions along both spatial dimensions. On the other hand, for studying topological properties via the charge pump, we consider a cylinder geometry in the presence of a flux $\theta$ threaded along the axis (see Fig. \ref{figure_cylinder_lattice}). Assuming that the zigzag edge of the cylinder is oriented along the $x$ axis and the cylinder axis is along the $y$ axis, we choose a gauge such that all nearest-neighbour hopping amplitudes crossing the line $x = 0$ acquire an additional phase factor $\exp\left(\pm i\theta\right)$ for hopping to the right (left) (see Fig. \ref{figure_lattice_2D}). This is equivalent to twisted boundary conditions along $x$ and open boundary conditions along $y$.\par\noindent
		The noninteracting time-periodically driven system has been shown to host a Chern insulator phase at high driving frequencies and an AFTI phase at intermediate driving frequencies \cite{dutta2024anomalous, zheng2024floquet}. While the former is characterized by topologically nontrivial edge modes appearing inside the bandgap for nonvanishing bulk Chern numbers, the latter is distinguished by edge modes appearing across all bandgaps despite vanishing bulk Chern numbers. A major motivation of our study will be to investigate the emergent phases and their stabilities in these two different regimes of driving frequency for the time-periodically driven interacting system.
	\section{Real-space Floquet Dynamical Mean Field Theory}
	\label{section_rfdmft}\noindent
		\noindent In the limit of infinite co-ordination number, the correlated lattice model can be mapped onto a set of self-consistent coupled Anderson impurity problems \cite{georges1996dynamical}. In order to probe the NESS of an inhomogeneous system, we use RFDMFT \cite{qin2018nonequilibrium, qin2017spectral}, leading to a set of coupled quantum impurity problems. Using the Floquet-Keldysh formalism \cite{tsuji2008correlated, aoki2014nonequilibrium, 2011theoretical, stefanucci2013nonequilibrium}, we express the nonequilibrium single particle Green's function in the lattice-site-Floquet basis,
		\begin{equation}
			\begin{split}
				\left[\pmb{G}\!\left(\omega\right)\right]_{\pmb{i}\pmb{j}, mn} = \mqty[\left[\pmb{G}^{R}\!\left(\omega\right)\right]_{\pmb{i}\pmb{j}, mn} & \left[\pmb{G}^{K}\!\left(\omega\right)\right]_{\pmb{i}\pmb{j}, mn}\\
				0 & \left[\pmb{G}^{R}\left(\omega\right)\right]^{\dagger}_{\pmb{i}\pmb{j}, mn}],
			\end{split}
		\end{equation}\noindent
		where $m,~n~\in\mathbb{Z}$ denote the photon sectors, $\omega\in[-\Omega / 2, \Omega / 2[$, while the boldfaces in $\pmb{G}^{R}$, the retarded Green's function component and in $\pmb{G}^{K}$, the Keldysh Green's function component, represent matrices. In Appendix \ref{appendix_nonequilibrium_greens_functions}, we present a brief review of nonequilibrium Green's functions as well as an analogue of the fluctuation-dissipation theorem in case of nonequilibrium time-periodically driven systems.\par\noindent
		To study the NESS of the time-periodically driven system, we also couple each lattice-site to a noninteracting fermionic heat bath in equilibrium at finite temperature $T$ as in previous studies \cite{tsuji2008correlated, aoki2014nonequilibrium, qin2017spectral, qin2018charge, qin2018nonequilibrium}. The total Hamiltonian including the system and the bath, $\mathcal{H}_{T}\!\left(t\right)$, is given by:
		\begin{equation}
				\mathcal{H}_{T}\!\left(t\right) = \mathcal{H}_{S}\!\left(t\right) + \mathcal{H}_{b} + \mathcal{H}_{S,b},
		\end{equation}\par\noindent
		where $\mathcal{H}_{b}$ represents the bath while $\mathcal{H}_{S, b}$ represents the coupling between the system and the bath (the latter is assumed to be large compared to the system described by $\mathcal{H}_{S}\!\left(t\right)$). We consider B\"uttiker's heat-bath model \cite{buttiker1985small, buttiker1986role}
		\begin{equation}
			\begin{split}
				\mathcal{H}_{b} = & \sum\limits_{\pmb{i}}\sum\limits_{p}\epsilon_{b, p}\hat{b}_{\pmb{i}, p}^{\dagger}\hat{b}_{\pmb{i}, p}^{\vphantom{\dagger}}\\
				\mathcal{H}_{S,b} = & \sum\limits_{\bm{i}}\sum\limits_{p}V_{p}\!\left(\hat{b}_{\pmb{i}, p}^{\dagger}\hat{c}_{\bm{i}}^{\vphantom{\dagger}} + h.c.\right),
			\end{split}
		\end{equation}
		where $\hat{b}_{\pmb{i}, p}^{\vphantom{\dagger}}\!\left(\hat{b}_{\pmb{i}, p}^{\dagger}\right)$ annihilates (creates) a bath fermion at lattice-site $\pmb{i}$ in an internal state $p$, $\epsilon_{b, p}$ is the bath kinetic energy for the internal state $p$ and $V_{p}$ denotes the hybridization at each lattice-site between an itinerant fermion and a bath fermion in the internal state $p$. The bath is assumed to be in equilibrium at a finite temperature $T$ and the bath chemical potential, $\mu_{b}$, is determined such that there is no net flow of current between the bath and the system. We work with $T / J = 0.01$.\par\noindent
		In the next step, the bath degrees of freedom are integrated out. Since the Hamiltonian is quadratic in terms of the bath degrees of freedom, this integral can be computed analytically. As a result, the self-energy of the system now contains a bath self-energy contribution which is local in real space. The retarded component of the bath self-energy reads
		\begin{equation}
			\begin{split}
				\left[\pmb{\check{\Sigma}}_{b}^{R}\!\left(\omega\right)\right]_{\pmb{ij}, mn} = & \delta_{\pmb{ij}}\left[\pmb{\Sigma}_{b}^{R}\!\left(\omega\right)\right]_{\pmb{i}, mn}\\
				\left[\pmb{\Sigma}_{b}^{R}\!\left(\omega\right)\right]_{\pmb{i}, mn} = & \sum\limits_{p}\dfrac{V_{p}^{2}}{\omega + n\Omega + \mu_{b} - \epsilon_{b, p} + i\xi}\delta_{mn},
			\end{split}
		\end{equation}
		where $\xi\rightarrow0^{+}$ is a real positive infinitesimal quantity used for convergence and we set the reduced Planck's constant $\hbar = 1$. Using the Sokhotski-Plemelj theorem \cite{plemelj1908erganzungssatz}, we separate the real and the imaginary parts of the retarded self energy and define the spectral density for the bath $\Gamma\left(\omega + n\Omega\right)$
		\begin{equation}
			\begin{split}
				\left[\pmb{\Sigma}_{b}^{R}\!\left(\omega\right)\right]_{\pmb{i}, mn} = & -i\pi \sum\limits_{p}V_{p}^{2}\delta\!\left(\omega + n\Omega + \mu_{b} - \epsilon_{b, p}\right)\delta_{mn}\\
				& + \mathcal{P}\!\left(\sum\limits_{p}\dfrac{V_{p}^{2}}{\omega + n\Omega + \mu_{b} - \epsilon_{b, p}}\right)\delta_{mn}\\
				\Gamma\!\left(\omega + n\Omega\right) = & \pi \sum\limits_{p}V_{p}^{2}\delta\!\left(\omega + n\Omega + \mu_{b} - \epsilon_{b, p}\right),
			\end{split}
		\end{equation}
		where $\mathcal{P}$ denotes the principal value. We choose the bath chemical potential $\mu_{b}$ such that the principal value vanishes. As a result, the bath self-energy contribution to the retarded component becomes purely imaginary. Thus, there is no shift in the system chemical potential due to coupling to the bath and consequently, no net exchange of particles. Furthermore, we assume the bath spectral density to be constant, i.e., frequency-independent. The local retarded bath self-energy is then given by
		\begin{equation}
			\begin{split}
				\left[\pmb{\Sigma}_{b}^{R}\!\left(\omega\right)\right]_{\pmb{i}, mn} = & -i\Gamma\delta_{mn}.
			\end{split}
		\end{equation}\par\noindent
		The Keldysh component can be calculated from the retarded component, using the fluctuation-dissipation theorem \cite{stefanucci2013nonequilibrium}
		\begin{equation}
			\begin{split}
				\left[\pmb{\check{\Sigma}}_{b}^{K}\!\left(\omega\right)\right]_{\pmb{ij}, mn} = & F\!\left[\omega + \!\left(\dfrac{m + n}{2}\right)\Omega\right]\\
				& \times\left(\left[\pmb{\tilde{\Sigma}}_{b}^{R}\!\left(\omega\right)\right]_{\pmb{ij}, mn} - \left[\pmb{\check{\Sigma}}_{b}^{R}\!\left(\omega\right)\right]_{\pmb{ji}, nm}^{*}\right)\\
				\left[\pmb{\check{\Sigma}}_{b}^{K}\!\left(\omega\right)\right]_{\pmb{ij}, mn} = & \delta_{\pmb{i}\pmb{j}}\left[\pmb{\Sigma}_{b}^{K}\!\left(\omega\right)\right]_{\pmb{i}, mn} \\
				\left[\pmb{\Sigma}_{b}^{K}\!\left(\omega\right)\right]_{\pmb{i}, mn} = & -2i\Gamma F\!\left(\omega + n\Omega\right)\delta_{mn}\\
				F\!\left(\omega\right) = & \tanh\!\left(\dfrac{\omega}{2k_{B}T}\right),
			\end{split}
		\end{equation}
		where $k_{B}$ is the Boltzmann constant and we set $k_{B} = 1$. We use $\Gamma / J = 0.005$ for all calculations in this paper. \par\noindent
		Subsequently, we define the Floquet Hamiltonian and express it in the lattice-site-Floquet basis 
		\begin{equation}
			\begin{split}
				\mathcal{H}_{0}\!\left(t\right) = & \sum\limits_{\pmb{i}\pmb{j}}\left[\pmb{\mathcal{H}}_{0}\!\left(t\right)\right]_{\pmb{i}\pmb{j}}c^{\dagger}_{\pmb{i}}c^{\vphantom{\dagger}}_{\pmb{j}}\\
				\left[\pmb{\mathcal{H}}_{0}\right]_{\pmb{i}\pmb{j}, mn} = & \dfrac{1}{\tau}\int\limits_{\left<\tau\right>}dt \left[\pmb{\mathcal{H}}_{0}\!\left(t\right)\right]_{\pmb{i}\pmb{j}}\exp\!\left[i\!\left(m - n\right)\Omega t\right]\\
				&  - \delta_{\pmb{i}\pmb{j}}\delta_{mn}m\Omega,
			\end{split}
		\end{equation}
		where $\left<\tau\right>$ refers to an integral over the period $\tau$. Since the Hamiltonian is periodic with respect to time, the integral can be performed over any interval of real time $t$ such that the final and initial points differ by $\tau$. Furthermore, since $\mathcal{H}_{0}\left(t\right)$ is Hermitian, the Floquet Hamiltonian becomes Hermitian as well.
		\begin{equation}
			\begin{split}
				\mathcal{H}_{0}\!\left(t\right) = & \mathcal{H}_{0}^{\dagger}\!\left(t\right)\\
				\left[\pmb{\mathcal{H}}_{0}\!\left(t\right)\right]_{\pmb{i}\pmb{j}} = & \left[\pmb{\mathcal{H}}_{0}\!\left(t\right)\right]_{\pmb{j}\pmb{i}}^{*}\\
				\left[\pmb{\mathcal{H}}_{0}\right]_{\pmb{i}\pmb{j}, mn} = & \left[\pmb{\mathcal{H}}_{0}\right]_{\pmb{j}\pmb{i}, nm}^{*}.
			\end{split}
		\end{equation}
		We then define $\pmb{G}_{0}$, the noninteracting Green's function via\par\noindent
		\begin{equation}
			\begin{split}
				\left[\pmb{G}^{-1}_{0}\!\left(\omega\right)\right]^{R}_{\pmb{i}\pmb{j}, mn} = & \left(\omega + \mu + i\eta\right)\delta_{\pmb{ij}}\delta_{mn} - \left[\pmb{\mathcal{H}}_{0}\right]_{\pmb{i}\pmb{j}, mn},
			\end{split}
		\end{equation}\noindent
		where $\mu = U / 2$ is the chemical potential at half-filling and $\eta\rightarrow0$ is a positive real convergence factor, which can be ignored in the presence of the thermal bath. Furthermore, we set the noninteracting Keldysh component to be zero, as is the case for a time-periodically driven system in the presence of a thermal bath \cite{tsuji2009photoinduced, aoki2014nonequilibrium}.
		\begin{equation}
			\begin{split}
				\left[\pmb{G}_{0}^{-1}\!\left(\omega\right)\right]_{\pmb{ij}, mn}^{K} = & 0.
			\end{split}
		\end{equation}\par\noindent
		We express $\mathcal{H}_{0}\left(t\right)$ using the modified Bessel functions of the first kind $I_{l}, ~l~\in~\mathbb{Z}$ \cite{arfken2011mathematical} which yields
		\begin{equation}
			\begin{split}
				\left[\pmb{\mathcal{H}}_{0}\right]_{\pmb{i}\pmb{j}, mn} = & \sum\limits_{\gamma = 1}^{3}\delta_{\pmb{j}, \pmb{i} + \pmb{\beta}_{\pmb{i}\gamma}}\left[J I_{n - m}\!\left(B\right)e^{i\!\left(n - m\right)\phi_{\gamma}}\right]\\
				& - m\Omega\delta_{\pmb{i}\pmb{j}}\delta_{mn}.
			\end{split}
		\end{equation}\noindent
		We assume a local but site-dependent lattice self-energy,
		\begin{equation}
			\begin{split}
				\left[\pmb{\check{\Sigma}}\!\left(\omega\right)\right]_{\pmb{i}\pmb{j}, mn} = &  \delta_{\pmb{i}\pmb{j}}\left[\pmb{\Sigma}\!\left(\omega\right)\right]_{\pmb{i}, mn}.
			\end{split}
		\end{equation}\noindent
		In our calculations, we restrict the photon sectors to $\left|m\right|, \left|n\right|\leq2$ and discretize the real frequency $\omega\in[-\Omega / 2, \Omega / 2[$ into $n_{f}$ = 1024 grid points. For the cylinder geometry, we perform the same computations with different boundary conditions(Fig. \ref{figure_cylinder_lattice}), by varying the threaded flux $\theta$ over an interval of $[-\pi, \pi[$, discretized into $n_{\theta} = 41$ grid points. For each value of the driving frequency $\Omega$ we start our calculations for a very small interaction strength relative to the driving frequency (i.e., a very small $U / \Omega$) and then increase $U / \Omega$ in small steps. For $U / \Omega = 0$, the Hamiltonian becomes noninteracting, leading to $\left[\pmb{\Sigma}\!\left(\omega\right)\right]_{\pmb{i}, mn} = 0$. We initialize the lattice self-energy in case of the bulk calculations for every $U / \Omega \neq 0$ with the converged bulk self-energy obtained for the value of the interaction strength relative to the driving frequency nearest to, but smaller than, $U / \Omega$ in magnitude. For the charge pump calculations, we initialize the lattice self-energy for every value of $U / \Omega \neq 0$ and $\theta$ with the converged cylinder geometry self-energy obtained for zero flux and the value of the interaction strength relative to the driving frequency nearest to, but smaller than, $U / \Omega$ in magnitude.\par\noindent
		For each driving frequency $\Omega$, interaction strength $U / \Omega$ (and threaded flux $\theta / 2\pi$ in the case of the cylinder geometry), we begin the RFDMFT iteration for each $\omega\in[- \Omega / 2, \Omega / 2[$ by calculating $\pmb{G}$ using the lattice Dyson equation,
		\begin{equation}
			\begin{split}
				\left[\pmb{G}^{-1}\!\left(\omega\right)\right]_{\pmb{i}\pmb{j}, mn} = & \left[\pmb{G}_{0}^{-1}\!\left(\omega\right)\right]_{\pmb{i}\pmb{j}, mn} - \delta_{\pmb{i}\pmb{j}}\!\left[\pmb{\Sigma}\!\left(\omega\right)\right]_{\pmb{i}, mn} \\
				& - \delta_{\pmb{i}\pmb{j}}\!\left[\pmb{\Sigma}_{b}\!\left(\omega\right)\right]_{\pmb{i}, mn}.
			\end{split}
		\end{equation}\noindent
		We denote all quantities belonging to the Anderson impurity model at a lattice-site $\pmb{i}$ using the superscript index $\!\left(\pmb{i}\right)$.\par\noindent
		We calculate for each lattice-site $\pmb{i}$ the Weiss function, which incorporates the influence of all other sites on $\pmb{i}$,
		\begin{equation}
			\left[\pmb{\mathcal{G}}^{\left(\pmb{i}\right)}\!\left(\omega\right)\right]_{mn}^{-1} = \left[\pmb{G}\!\left(\omega\right)\right]^{-1}_{\pmb{ii}, mn} + \left[\pmb{\Sigma}\!\left(\omega\right)\right]_{\pmb{i}, mn}.
		\end{equation}\noindent
		In the limit of infinite co-ordination number, the local Green's functions $\pmb{G}^{\left(\pmb{i}\right)}$ for the Falicov-Kimball model can be calculated exactly using the Weiss function $\pmb{\mathcal{G}}^{\left(\pmb{i}\right)}$ \cite{brandt1989thermodynamics, brandt1990thermodynamics, brandt1991free, eckstein2009nonequilibrium}, as follows
		\begin{equation}
			\begin{split}
				\left[\pmb{G}^{\left(\pmb{i}\right)}\!\left(\omega\right)\right]_{mn} = &  \left(1 - w_{\pmb{i}}\right)\left[\pmb{\mathcal{G}}^{\left(\pmb{i}\right)}\!\left(\omega\right)\right]_{mn} \\
				& + w_{\pmb{i}}\left[\left[\pmb{\mathcal{G}}^{\left(\pmb{i}\right)}\!\left(\omega\right)\right]^{-1} - U\pmb{\mathds{1}}\right]_{mn}^{-1},
			\end{split}
			\label{equation_impurity_solver}
		\end{equation}\noindent
		where $w_{\pmb{i}}$ is the probability of impurity site $\pmb{i}$ being occupied by an immobile fermion and $\pmb{\mathds{1}}_{mn} = \delta_{mn}$ is the identity matrix. We set $w_{\pmb{i}} = 0.5\,\forall\,\pmb{i}$, as in the homogeneous phase that emerges in the equilibrium model above the transition temperature \cite{brandt1989thermodynamics, brandt1990thermodynamics, brandt1991free}. We then obtain for each lattice-site $\pmb{i}$ the updated local self-energy from the local Dyson equation
		\begin{equation}
			\left[\pmb{\Sigma}^{\left(\pmb{i}\right)}\!\left(\omega\right)\right]_{mn} = \left[\pmb{\mathcal{G}}^{\left(\pmb{i}\right)}\!\left(\omega\right)\right]_{mn}^{-1} - \left[\pmb{G}^{\left(\pmb{i}\right)}\!\left(\omega\right)\right]_{mn}^{-1}.
		\end{equation}\noindent
		At the end of every RFDMFT iteration, we compute the quantity
		\begin{equation}
			\chi\!\left(\omega\right) = \max\limits_{\pmb{i}, m, n}\left|\left[\pmb{\Sigma}^{\left(\pmb{i}\right)}\!\left(\omega\right)\right]_{mn} - \left[\pmb{\Sigma}\!\left(\omega\right)\right]_{\pmb{i}, mn}\right|.
		\end{equation}\noindent
		Considering a tolerance $\epsilon$, if $\chi\!\left(\omega\right) < \epsilon$, the algorithm has converged. If $\chi\!\left(\omega\right) \geq \epsilon$, $\left[\pmb{\Sigma}\!\left(\omega\right)\right]_{\pmb{i}, mn}$ is updated. Subsequently, using the updated lattice self-energy, a new RFDMFT iteration is initated. \par\noindent
		Due to the exact impurity solver for the Falicov-Kimball model (Eq. \eqref{equation_impurity_solver}), the RFDMFT calculation for each frequency component of the self energy can be performed independently, enabling parallelization over different $\omega$ and different $\theta$. Updating the lattice self-energies using the Newton-Rhapson method~\cite{ypma1995historical} where the local self-energies are mixed with the current lattice self-energies linearly, we obtain converged results up to a tolerance of $10^{-5}$ in the regime of high driving frequency (in our case, $\Omega / J = 18$) within 500 iterations, across all $\omega$ and $\theta$. However, for intermediate values of the driving frequency (in our case, $\Omega / J = 8.7$), even 20000 iterations are not sufficient to achieve convergence up to the same tolerance across all $\omega$ and $\theta$. The problem is particularly severe near the edges of the Floquet Brillouin zone, which seems analogous to the convergence issues near quasimomentum band edges in \cite{freericks2003exact}.\par\noindent 
		Since good convergence is essential to obtain faithful results, we address this problem by using the Broyden solver for updating the local lattice self-energies, as detailed in Appendix \ref{appendix_broyden_solver}. This solver is a quasi-Newton-Rhapson method that improves convergence for iteratively solving systems of nonlinear equations and has been implemented and benchmarked for equilibrium DMFT~\cite{broyden1965class, dederichs1983self, vanderbilt1984total, srivastava1984broyden, vzitko2009convergence}. Such a scheme takes as input the current local self-energy $\left[\pmb{\Sigma}^{\left(\pmb{i}\right)}\!\left(\omega\right)\right]_{mn}$ along with the lattice self-energies from all previous iterations (upto a desired cutoff, taken to be 50 in our case). However, for some parameter values in the cylinder geometry the self-energy still does not converge to the desired tolerance.
	\section{Bulk Spectrum}
	\label{section_bulk_spectrum}\noindent
		\noindent For studying the bulk spectrum, we impose periodic boundary conditions along both spatial dimensions. We then employ RFDMFT (Section \ref{section_rfdmft}) to obtain the nonequilibrium Green's functions expressed in the lattice-site-Floquet basis. Using these nonequilibrium Green's functions, we calculate the bulk spectral function and occupied density of states,
		\begin{equation}
			\begin{split}
				\mathcal{A}\!\left(\omega'\right) = & -\dfrac{1}{\pi N_{s}}\sum\limits_{\pmb{i}}\operatorname{Im}\left[\pmb{G}^{R}\!\left(\omega\right)\right]_{\pmb{i}\pmb{i}, nn}\\
				\mathcal{N}\!\left(\omega'\right) = & \dfrac{1}{4\pi N_{s}}\sum\limits_{\pmb{i}}\left[\operatorname{Im}\left[\pmb{G}^{K}\!\left(\omega\right)\right]_{\pmb{i}\pmb{i}, nn} \right.\\
				& \left. - 2\operatorname{Im}\left[\pmb{G}^{R}\!\left(\omega\right)\right]_{\pmb{i}\pmb{i}, nn}\right],
			\end{split}
		\end{equation}\noindent
		where $N_{s}$ is the number of lattice-sites and $\omega' \equiv \omega + n\Omega$ is in the full range of the real frequency spectrum.\par\noindent
		\begin{figure}[!t]
			\centering
			\includegraphics[width=\linewidth, keepaspectratio]{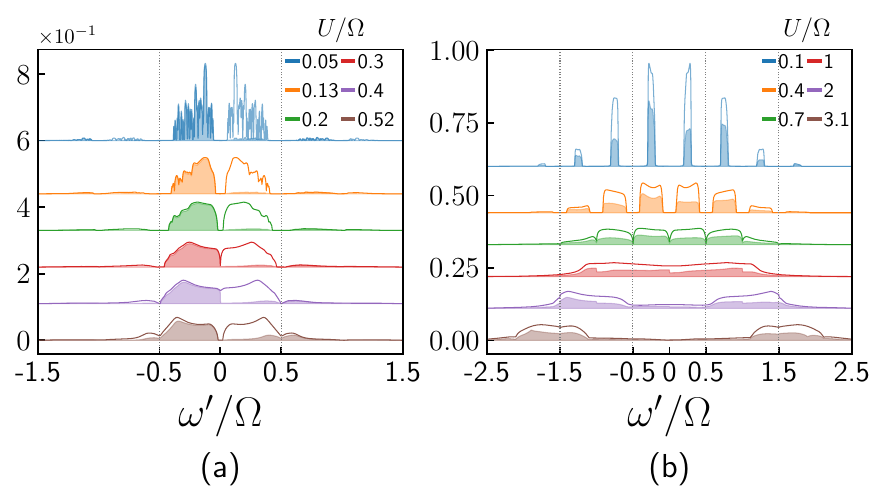}
			\caption{Spectral function $\mathcal{A}\!\left(\omega'\right)$ (solid lines) and occupied density of states $\mathcal{N}\!\left(\omega'\right)$ (solid fill) as functions of real frequency $\omega'$, calculated on a $16 \times 16$ lattice for 1024 values of the real frequency $\omega^{\prime}$ at a) $\Omega / J$ = 18, b) $\Omega / J$ = 8.7. For each $\Omega$, the spectrum is shifted vertically for different $U$.}
			\label{figure_bulk_dos}
		\end{figure}\par\noindent
		In Fig. \ref{figure_bulk_dos}, the bulk spectral function is shown as a function of the real frequency $\omega ' / \Omega$ in line plots for various interaction strength $U$ while the occupied density of states is denoted by the shaded region, for two regimes with different driving frequency $\Omega$. For each $\Omega$, the spectra for different interaction strengths $U$ are represented with different vertical shifts.\par\noindent
		For $\Omega / J = 18$, the occupied density of states is small in the upper Mott band, manifested by $\mathcal{N}\sim0$ for large $\omega$ and $n$. This resembles the equilibrium Haldane-Falicov-Kimball model occupied density of states. With increasing $U / \Omega$, the Hubbard bands shift closer until they merge at the Mott transition. For $U / \Omega$ above this critical value, a finite band gap exists between the two bands.\par\noindent
		For $\Omega / J = 8.7$, the occupied density of states is distributed across all bands. Thus, there can be no effective equilibrium description of the system at this intermediate driven frequency, which is comparable to the energy scales of the system. With increasing $U / \Omega$, the bands begin to coalesce. At the Mott transition, they merge. For $U / \Omega$ above this critical value, two bands exist with a finite band gap.
	\section{Edge Spectrum}
	\label{section_edge_spectrum}\noindent
		\noindent For studying the charge pump and edge states, we consider a cylinder geometry in the presence of a flux threaded along the axis (see Fig. \ref{figure_cylinder_lattice}). We then employ RFDMFT (Section \ref{section_rfdmft}) to obtain the nonequilibrium Green's functions expressed in the lattice-site-Floquet basis. Using these nonequilibrium Green's functions, we calculate for every value of the flux $\theta$ the spectral function\par\noindent
		\begin{equation}
			\begin{split}
				\mathcal{A}\!\left(\omega', \theta\right) = & -\dfrac{1}{\pi N_{s}}\sum\limits_{\pmb{i}}\operatorname{Im}\left[\pmb{G}^{R}\!\left(\omega, \theta\right)\right]_{\pmb{i}\pmb{i}, nn}.
			\end{split}
		\end{equation}\par
\begin{figure*}[!t]
			\centering
			\includegraphics[width=\linewidth, keepaspectratio]{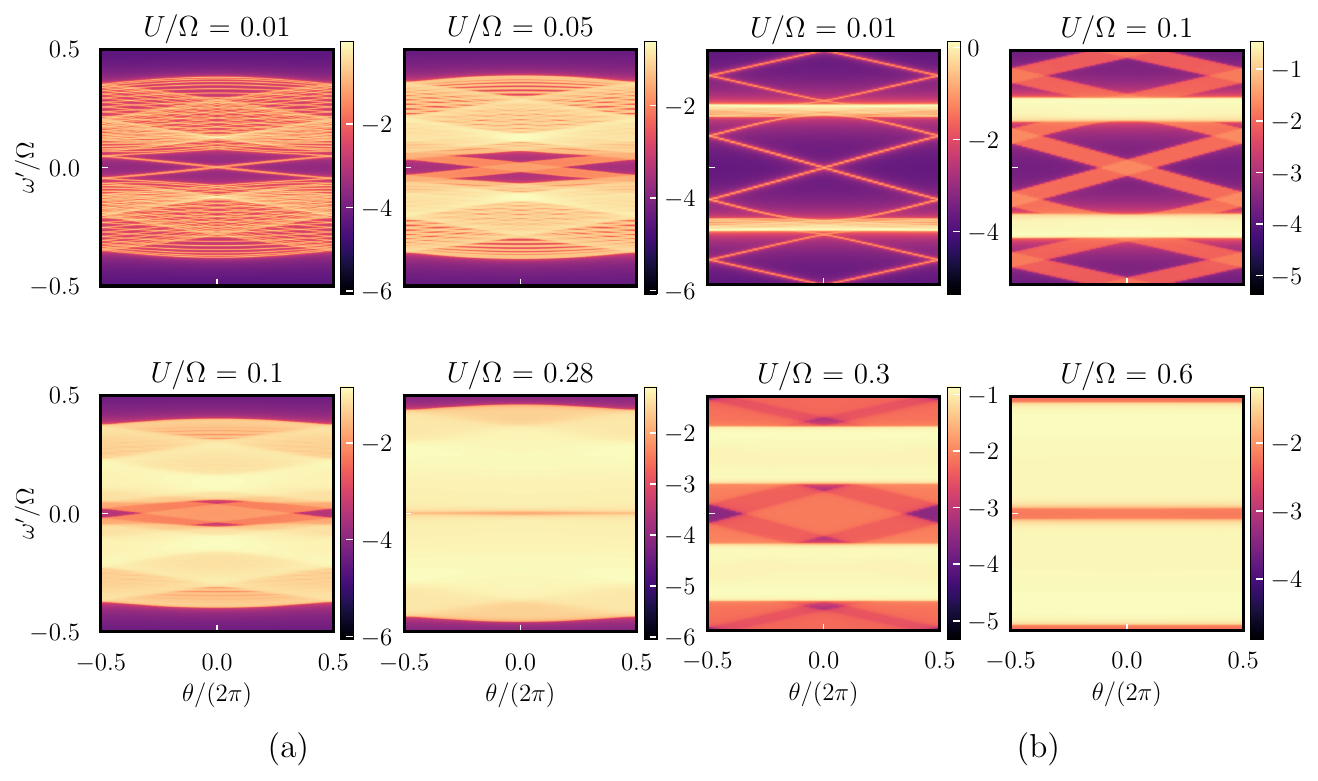}
			\caption{Spectral function $\mathcal{A}\!\left(\omega', \theta\right)$ for the cylinder geometry as a function of threaded flux $\theta$ and real frequency $\omega'$, on an $8 \times 16$ lattice, for 1024 values of the real frequency $\omega^{\prime}$ and 21 values of the flux $\theta$ at a) $\Omega / J$ = 18, b) $\Omega / J$ = 8.7.}
			\label{figure_edge_spectrum}
		\end{figure*}

\noindent
		In Fig. \ref{figure_edge_spectrum} the spectral function is shown in a false colour plot as a function of the real frequency $\omega' = n \Omega + \omega$ over the full range of the spectrum and the flux $\theta$ for various interaction strengths $U$ for the two regimes with different driving frequency $\Omega$ as investigated for the bulk spectrum (Section \ref{section_bulk_spectrum}). Both regimes represent nontrivial topological phases for $U / \Omega \sim0$. However, while for $\Omega / J$ = 18 the edge modes exist only in the bandgaps around $\omega^{\prime} = n\Omega, n\in\mathbb{Z}$, they exist within all bandgaps for $\Omega / J$ = 8.7, indicating an anomalous topological phase \cite{rudner2013anomalous, kitagawa2010topological, zheng2024floquet}. In both regimes, with increasing $U / \Omega$, the spectral function of the edge modes begins to broaden and becomes less sensitive to changes in $\theta$. At the Mott transition, the spectrum becomes flat with respect to $\theta$ and no edge modes are visible anymore.
		\par\noindent
	\section{Charge Pump}
	\label{section_charge_pump}\noindent
		To study the topological phase transitions of the system, we segment the cylinder geometry into the lower (L) and the upper (U) halves according to the axial length (Fig. \ref{figure_cylinder_lattice}). We then calculate the densities in the two halves of the cylinder\par\noindent
		\begin{equation}
			\begin{split}
				Q^{L\left(U\right)}\!\left(\theta\right) = & \sum\limits_{\bm{i}\in L\left(U\right)}\sum\limits_{n}\int\limits_{-\Omega / 2}^{\Omega / 2}\dfrac{d\omega}{4\pi}\left[\operatorname{Im}\left[\pmb{G}^{K}\!\left(\omega, \theta\right)\right]_{\pmb{i}\pmb{i}, nn} \right. \\
				& \left. - 2\operatorname{Im}\left[\pmb{G}^{R}\!\left(\omega, \theta\right)\right]_{\pmb{i}\pmb{i}, nn}\right].
			\end{split}
		\end{equation}
		Subsequently, we calculate for each flux $\theta$ the difference $Q\!\left(\theta\right)$ between these densities. In the case of a topologically nontrivial phase, a jump discontinuity emerges in $Q\!\left(\theta\right)$ as the threaded flux $\theta$ is varied over an interval of length $2\pi$~\cite{laughlin1981quantized}. Hence, we use the extrema in $Q\!\left(\theta\right)$ with respect to the threaded flux $\theta$ to compute the pumped charge. \par\noindent
		\begin{equation}
			\begin{split}
				Q\!\left(\theta\right) = & Q^{U}\!\left(\theta\right) - Q^{L}\!\left(\theta\right)\\
				P = & \dfrac{\left[\operatorname{max}\limits_{\theta}Q\!\left(\theta\right) - \operatorname{min}\limits_{\theta}Q\!\left(\theta\right)\right]}{2}.
			\end{split}
			\label{equation_topological_charge}
		\end{equation}\par\noindent
		\begin{figure}[!hb]
			\centering
			\includegraphics[width=\linewidth, keepaspectratio]{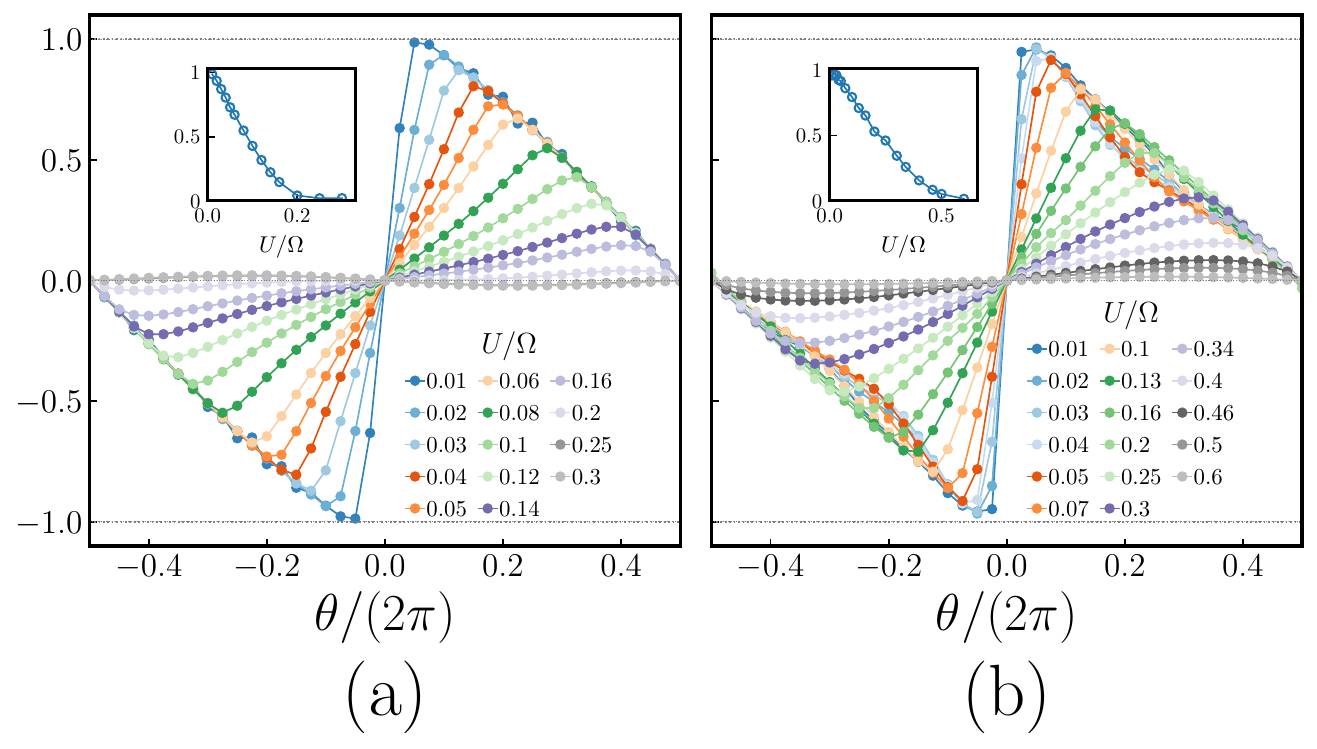}
			\caption{Difference in charge between the lower and upper halves of the cylinder, $Q$ and the pumped charge $P$ (inset) as functions of the threaded flux $\theta$, calculated on an $8 \times 16$ lattice for 1024 values of the real frequency $\omega^{\prime}$ and 21 values of the flux $\theta$ at a) $\Omega / J$ = 18, b) $\Omega / J$ = 8.7.}
			\label{figure_edge_charge}
		\end{figure}\par\noindent
		Fig. \ref{figure_edge_charge} shows the difference in charge $Q$ between the two halves of the cylinder as a function of the flux $\theta$ for various interaction strengths $U$ for the two regimes with different driving frequency $\Omega$. In both driving regimes in the presence of a finite interaction strength $U$, $P$ is no longer quantized unlike for a Laughlin pump in a closed system in the absence of the bath~\cite{laughlin1981quantized}. With increasing $U$, $P$ keeps decreasing until it becomes very small at and beyond the Mott transition. This signifies a shift to the topologically trivial Mott insulating phase, occuring for $U / \Omega \gtrsim 0.3$ for both values of $\Omega / J$.
	\section{Local Density of States}
	\label{section_ldos}\noindent
		We also investigate the local density of states at zero frequency and zero flux\par\noindent
		\begin{equation}
			\begin{split}
				\mathcal{A}^{\left(0\right)}_{\left(i_{x}, i_{y}\right)} = -\dfrac{1}{\pi}\operatorname{Im}\left[\pmb{G}^{R}\!\left(\omega = 0, \theta = 0\right)\right]_{\left(i_{x}, i_{y}\right)\left(i_{x}, i_{y}\right), 00}.
			\end{split}
		\end{equation}\par
\begin{figure}[!h]
			\centering
			\includegraphics[width=\linewidth, keepaspectratio]{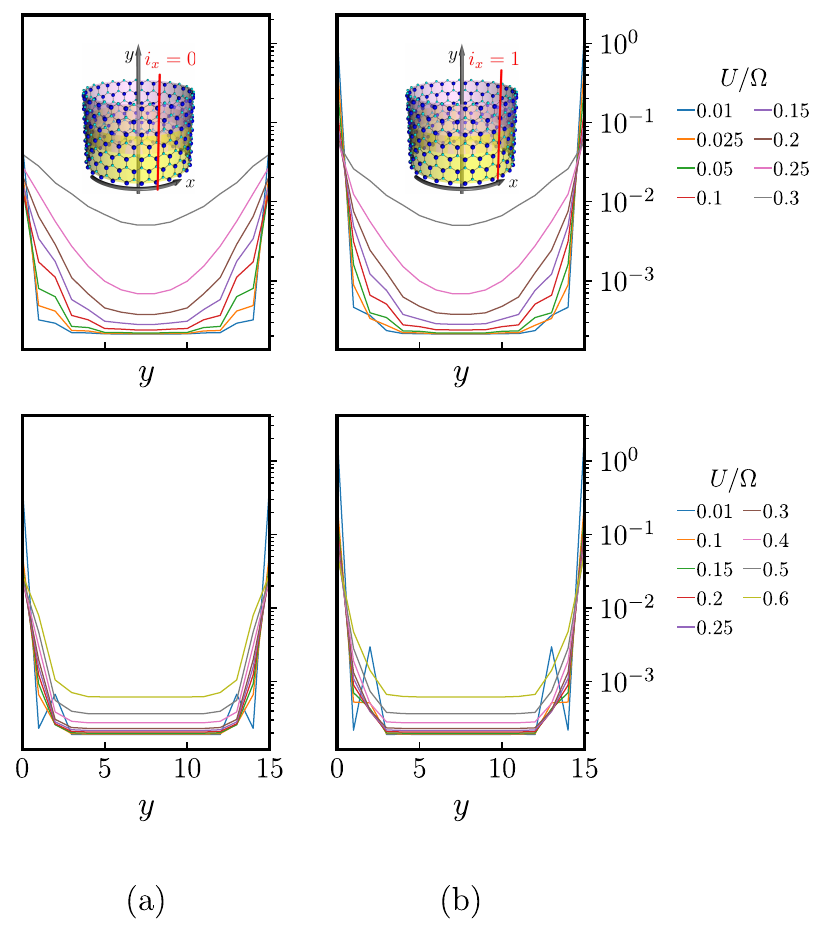}
			\caption{Variation of the local density of states at $\omega / \Omega = 0$, $\theta = 0$ along the open ($y$) direction of the cylinder at different values of $U / \Omega$, calculated on an $8 \times 16$ lattice, at a vertical cut along $i_{x}$ = 0 (left panels) and $i_{x}$ = 1 (right panels) at a) $\Omega / J$ = 18, b) $\Omega / J$ = 8.7. The insets in the top panels represent the cuts $i_{x} = 0$ and $i_{x} = 1$ for the left and right panels respectively.}
			\label{figure_ldos}
		\end{figure}

\noindent
		We investigate the dependence of $\mathcal{A}^{\left(0\right)}$ on $i_{y}$ for different $i_{x}$ in Fig. \ref{figure_ldos}. For both values of the driving frequency $\Omega$, the local density of states features a sharp drop from the edge towards the bulk for low values of the interaction strength $U$. However, as the interaction strength increases, there is a significant hybridzation into the bulk.\par\noindent
		\par\noindent
	\section{Energy Dissipation}
	\label{section_open_system}\noindent
		We investigate the dissipation of energy into the bath. Assuming a steady state, the dissipation becomes\par\noindent
		\begin{equation}
			\begin{split}
				I = & \Gamma\sum\limits_{\pmb{i}}\sum\limits_{n\in\mathbb{Z}}\int\limits_{-\Omega / 2}^{\Omega / 2}\dfrac{d\omega}{2\pi N_{s}}\left(\omega + n\Omega\right)\left[\operatorname{Im}\left[\pmb{G}^{K}\left(\omega\right)\right]_{\pmb{i}\pmb{i}, nn} \right.\\
				& \left. - 2F\left(\omega + n\Omega\right)\operatorname{Im}\left[\pmb{G}^{R}\left(\omega\right)\right]_{\pmb{i}\pmb{i}, nn}\right].
			\end{split}
		\end{equation}\par\noindent
		\begin{figure}[!ht]
			\centering
			\includegraphics[width=\linewidth, keepaspectratio]{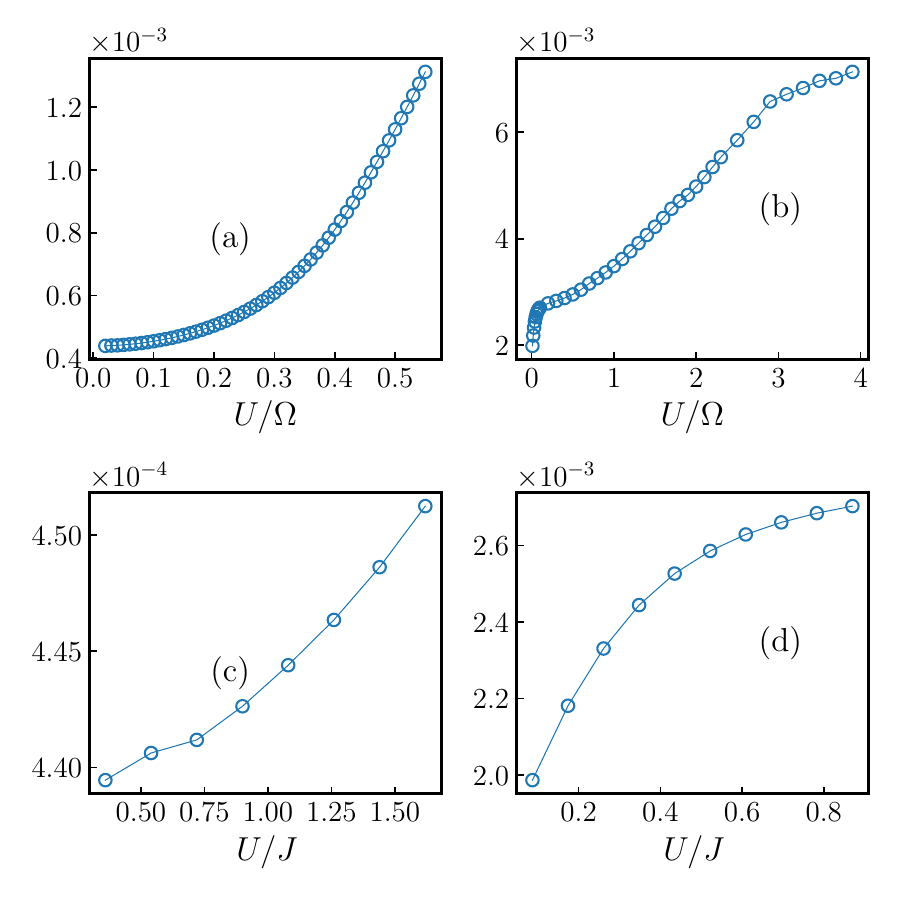}
				\caption{Bulk energy dissipation rate into the bath as a function of $U / \Omega$ for an $8 \times 16$ lattice at a) $\Omega / J$ = 18, b) $\Omega / J$ = 8.7. We further show the variation of this bulk dissipation rate as a function of $U / J$ for low interaction strength, in c) for $\Omega / J$ = 18 and in d) for $\Omega / J$ = 8.7.}
			\label{figure_diss}
		\end{figure}\par\noindent
		From Fig. \ref{figure_diss}, we observe that the dissipation increases monotonically with increasing interaction strength in the two driving regimes, $\Omega / J = 18$ for the Haldane phase and $\Omega / J = 8.7$ for the anomalous phase. This indicates interaction-induced Floquet heating, with many-body scattering processes leading to absorption of energy from the time-periodic drive~\cite{d2014long, lazarides2014equilibrium, bukov2015prethermal, mallayya2019heating}.\par\noindent
		The dissipation is suppressed in the Haldane phase while it is enhanced in the anomalous phase. In order to investigate the differences in dissipation between these two driving regimes further, we plot the dissipation for low values of $U$ in Fig. \ref{figure_diss}c) and d). For $\Omega / J = 18$, the dissipation exhibits a convex functional dependence on $U$, characteristic of off-resonant heating~\cite{mallayya2019heating}. Conversely for $\Omega / J = 8.7$, the dissipation with increasing $U$ exhibits concavity, indicating a high sensitivity to $U$ even in the weakly interacting limit $U \ll J$. This suggests that lower driving frequencies bring the many-body spectrum closer to many-body Floquet resonances, allowing even weak interactions to effectively mediate energy transfer from the time-periodic drive.
	\section{Conclusion}
	\label{section_conclusion}\noindent
		We have studied the spectrum, charge pump and energy dissipation into the bath for fermions in a time-periodically driven Haldane Falicov-Kimball model coupled to a noninteracting equilibrium finite-temperature fermionic bath. We have considered the regimes of large as well as intermediate driving frequencies, to investigate phases which can and cannot be described by an effective equilibrium Hamiltonian, respectively. We find that increasing the interaction strength leads to increasing broadening of the spectral function of the edge modes. This is accompanied by a decrease of the pumped charge observed in a Laughlin pump setup. At the Mott transition, the pumped charge becomes very small. We also calculate the local density of states at zero frequency and zero flux for the Laughlin pump and show that the corresponding edge mode increasingly hybridizes with the bulk for increasing interaction strengths. \par\noindent
		Due to the presence of a heat bath, the pumped charge is not quantized even for small interaction strengths. We verify this statement by probing the energy dissipation from the system to the bath. While the dissipation increases with increasing interaction strength for all driving frequency regimes investigated, the qualitative behaviour of the increase differs strongly.\par\noindent
		The usage of the Broyden solver to update the self-energy has led to faster convergence in RFDMFT, especially in the anomalous driving regime. The flexibility of using RFDMFT with the Broyden solver as well as the resultant faster convergence allow for application to a variety of experimentally realistic models. In the future, we will investigate the Haldane-Hubbard model where both spin states are mobile. Another interesting problem for future study would be bath engineering to optimize the energy dissipation and preserve the nontrivial topology of the phases. A third intriguing aspect could be investigated via the addition of quenched disorder, which would lead to the interplay between localization due to interaction and disorder.
	\section{Acknowledgement}\noindent
		This work was supported by the Deutsche Forschungsgemeinschaft (DFG, German Research Foundation) under Project No. 277974659 via Research Unit FOR 2414.The authors gratefully acknowledge the computing time provided to them at the NHR Center NHR@SW at Goethe University Frankfurt. This is funded by the Federal Ministry of Education and Research, and the state governments participating on the basis of the resolutions of the GWK for national high performance computing at universities (www.nhr-verein.de/unsere-partner). The authors gratefully acknowledge the Gauss Centre for Supercomputing e.V. (www.gauss-centre.eu) for funding this project by providing computing time through the John von Neumann Institute for Computing (NIC) on the GCS Supercomputer JUWELS at J\"ulich Supercomputing Centre (JSC).
	\appendix
	\section{Nonequilibrium Green's Functions}
	\label{appendix_nonequilibrium_greens_functions}\noindent
		We first define the retarded, Keldysh, lesser and greater Green's functions in the time domain, working with the Schr\"odinger picture as:
		\begin{equation}
			\begin{split}
				G_{\pmb{ij}}^{R}\!\left(t_{1}, t_{2}\right) = & -i\theta\!\left(t_{1} - t_{2}\right)\left<\left[\hat{c}_{\pmb{i}}^{\vphantom{\dagger}}\!\left(t_{1}\right), \hat{c}_{\pmb{j}}^{\dagger}\!\left(t_{2}\right)\right]_{\mp}\right>\\
				G_{\pmb{ij}}^{K}\!\left(t_{1}, t_{2}\right) = & -i\left<\left[\hat{c}_{\pmb{i}}^{\vphantom{\dagger}}\!\left(t_{1}\right), \hat{c}_{\pmb{j}}^{\dagger}\!\left(t_{2}\right)\right]_{\pm}\right>\\
				G_{\pmb{ij}}^{<}\!\left(t_{1}, t_{2}\right) = & \pm i\left<\hat{c}_{\pmb{j}}^{\dagger}\!\left(t_{1}\right)\hat{c}_{\pmb{i}}^{\vphantom{\dagger}}\!\left(t_{2}\right)\right>\\
				G_{\pmb{ij}}^{>}\!\left(t_{1}, t_{2}\right) = & - i\left<\hat{c}_{\pmb{i}}^{\vphantom{\dagger}}\!\left(t_{1}\right)\hat{c}_{\pmb{j}}^{\dagger}\!\left(t_{2}\right)\right>\\
				\left[A, B\right]_{\xi} = & AB - \xi BA, \qquad\xi\in\left\{1, -1\right\}\\
				\theta\left(x\right) = & \left\{\begin{array}{cc}
										0 & x < 0 \\
										1 & x \geq 0
										\end{array}\right.,
			\end{split}
			\label{equation_nonequilibrium_greens_functions}
		\end{equation}\noindent
		where the upper (lower) sign corresponds to fermionic (bosonic) operators~\cite{stefanucci2013nonequilibrium} and $\theta$ is the Heaviside function.\par\noindent
		We now define the Wigner and the Floquet representations for each of the Green's function components:
		\begin{equation}
			\begin{split}
				\left[\pmb{G}\!\left(\omega\right)\right]_{\pmb{ij}, n} = & \dfrac{1}{\tau}\int\limits_{0}^{\tau} dt_{\text{avg}}\int\limits_{-\infty}^{\infty} dt_{\text{rel}}\, G_{\pmb{ij}}\!\left(t_{1}, t_{2}\right)e^{i\omega t_{\text{rel}} + in\Omega t_{avg}}\\
				t_{\text{rel}} = & t_{1} - t_{2}, \qquad t_{\text{avg}} =  \dfrac{t_{1} + t_{2}}{2}.
			\end{split}
			\label{equation_wigner_transform}
		\end{equation}\noindent
		For systems in equilibrium with time-independent Hamiltonians, there exist similar analogues to all of the components defined above, except for the Keldysh component. The latter is essential to account for the initial correlations in the system. However, in stark contrast to the nonequilibrium Green's functions, the equilibrium Green's functions do not depend on $t_{\text{avg}}$ \cite{stefanucci2013nonequilibrium}. \par\noindent
		The fluctuation-dissipation theorem, derived for static equilibrium systems, relates the lesser and greater components
		\begin{equation}
			\begin{split}
				\pmb{G}_{\pmb{ij}}^{>}\!\left(\omega\right) = & \pm\exp\!\left(\beta\omega\right)\pmb{G}_{\pmb{ij}}^{<}\!\left(\omega\right).
			\end{split}
		\end{equation}\noindent
		Since this equilibrium relation is derived via a Fourier transformation of only the relative time, i.e., the difference between the two time arguments, $t_{\text{rel}}$, we can apply the equilibrium relation independently within each photon sector $n\in\mathbb{Z}$. In other words,
		\begin{equation}
			\begin{split}
				\left[\pmb{G}^{>}\!\left(\omega\right)\right]_{\pmb{ij}, n} = & \pm\exp\!\left(\beta\omega\right)\!\left[\pmb{G}^{<}\!\left(\omega\right)\right]_{\pmb{ij}, n}.
			\end{split}
			\label{equation_lesser_and_greater_nonequilibrium}
		\end{equation}\noindent
		From Eq. \eqref{equation_nonequilibrium_greens_functions}, it then follows that the lesser and greater components can furthermore be expressed in terms of linear combinations of the Keldysh and the retarded components.
		\begin{equation}
			\begin{split}
				\left[\pmb{G}^{>}\!\left(\omega\right)\right]_{\pmb{ij}, n} = & \dfrac{1}{2}\left(\left[\pmb{G}^{K}\!\left(\omega\right)\right]_{\pmb{ij}, n} - \left[\pmb{G}^{R}\!\left(\omega\right)\right]_{\pmb{ij}, n} \right.\\
				& \left. + \left[\pmb{G}^{R}\!\left(\omega\right)\right]_{\pmb{ji}, -n}^{*}\right)\\
				\left[\pmb{G}^{<}\!\left(\omega\right)\right]_{\pmb{ij}, n} = & \dfrac{1}{2}\left(\left[\pmb{G}^{K}\!\left(\omega\right)\right]_{\pmb{ij}, n} + \left[\pmb{G}^{R}\!\left(\omega\right)\right]_{\pmb{ij}, n} \right.\\
				& \left. - \left[\pmb{G}^{R}\!\left(\omega\right)\right]_{\pmb{ji}, -n}^{*}\right).
			\end{split}
			\label{equation_lesser_and_greater_retarded_keldysh}
		\end{equation}\noindent
		Thus, from Eq. \eqref{equation_lesser_and_greater_nonequilibrium} and Eq. \eqref{equation_lesser_and_greater_retarded_keldysh}, we apply the fluctuation-dissipation theorem to obtain
		\begin{equation}
			\begin{split}
				\left[\pmb{G}^{K}\!\left(\omega\right)\right]_{\pmb{ij}, n} = & F\!\left(\omega + n\Omega\right)\\ & \left(\left[\pmb{G}^{R}\!\left(\omega\right)\right]_{\pmb{ij}, n} - \left[\pmb{G}^{R}\!\left(\omega\right)\right]_{\pmb{ji}, -n}^{*}\right)\\
				F\!\left(\omega\right) = & \left\{\mqty{\coth\!\left(\dfrac{\omega}{2T}\right) & \text{bosons}\\[8pt] \tanh\!\left(\dfrac{\omega}{2T}\right) & \text{fermions}.}\right.
			\end{split}
		\end{equation}\noindent
		We now define the Floquet transform to obtain the Green's functions in the lattice-site-Floquet basis as used throughout the main text
		\begin{equation}
			\begin{split}
				\left[\pmb{G}\!\left(\omega\right)\right]_{\pmb{ij}, mn} = & 	\dfrac{1}{\tau}\int\limits_{0}^{\tau} dt_{\text{avg}}\int\limits_{-\infty}^{\infty} dt_{\text{rel}}G_{\pmb{ij}}\!\left(t_{1}, t_{2}\right)\\
				&\times\exp\!\left[i\!\left(\omega + m\Omega\right)t_{1} - i\!\left(\omega + 	m\Omega\right)t_{2}\right]\\[8pt]		\left[\pmb{G}\!\left(\omega\right)\right]_{\pmb{ij}, mn} = & 	\left[\pmb{G}\!\left(\omega + \dfrac{m + n}{2}\Omega\right)\right]_{\pmb{ij}, m - n},	
			\end{split}
		\end{equation}\noindent
		where the right hand side of the last equation denotes the Wigner transform of the Green's function defined in Eq. \eqref{equation_wigner_transform}.\par\noindent
		Thus, expressing the fluctuation-dissipation theorem in the lattice-site-Floquet basis, we obtain
		\begin{equation}
			\begin{split}
				\left[\pmb{G}^{K}\!\left(\omega\right)\right]_{\pmb{ij}, mn} = & F\!\left[\omega + \!\left(\dfrac{m + n}{2}\right)\!\Omega\right]\\
				& \times\left(\left[\pmb{G}^{R}\!\left(\omega\right)\right]_{\pmb{ij}, mn} - \left[\pmb{G}^{R}\!\left(\omega\right)\right]_{\pmb{ji}, nm}^{*}\right).
			\end{split}
		\end{equation}
	\section{Broyden Solver}
	\label{appendix_broyden_solver}\noindent
		We first select for every iteration $m$,
		\begin{equation}
			\begin{split}
				\pmb{V}^{\left(m\right)} = & \pmb{\Sigma}^{\left(m\right)}\\
				\pmb{F}^{\left(m\right)} = & \left[\pmb{\Sigma}^{\left(\pmb{i}\right)}\right]^{\left(m\right)} - \left[\pmb{\Sigma}\!\left(\omega\right)\right]_{\pmb{i}}^{\left(m\right)}	\\
				\pmb{\Delta F}^{\left(m\right)} = & \dfrac{\pmb{F}^{\left(m + 1\right)} - \pmb{F}^{\left(m\right)}}{\left|\pmb{F}^{\left(m + 1\right)} - \pmb{F}^{\left(m\right)}\right|}\\
				\pmb{\Delta V}^{\left(m\right)} = & \dfrac{\pmb{V}^{\left(m + 1\right)} - \pmb{V}^{\left(m\right)}}{\left|\pmb{V}^{\left(m + 1\right)} - \pmb{V}^{\left(m\right)}\right|},
			\end{split}
		\end{equation}\par\noindent
		where the superscript $\left(m\right)$ denotes that the quantities represented have been obtained from the $m$\textsuperscript{th} iteration.\par\noindent
		We then use Broyden's method to update the solution, mixing the most recent $\pmb{V}^{\left(m\right)}$ with the previous solutions obtained, such that the function $\pmb{F}^{\left(m\right)}$ is minimized. In other words,\par\noindent
		\begin{equation}
			\begin{split}
				\pmb{V}^{\left(m + 1\right)} = & \pmb{V}^{\left(m\right)} + \alpha \pmb{F}^{\left(m\right)} - \sum\limits_{n = 1}^{m -1}\sum\limits_{k = 1}^{m -1}w_{n}w_{k}\pmb{c}_{k}^{\left(m\right)}\pmb{\beta}_{kn}^{\left(m\right)}\pmb{U}^{\left(n\right)}\\
				\pmb{c}_{k}^{\left(m\right)} = & \left[\pmb{\Delta F}^{\left(k\right)}\right]^{\dagger}\pmb{F}^{\left(m\right)}\\
				\pmb{U}^{\left(n\right)} = & \alpha\pmb{\Delta F}^{\left(n\right)} + \pmb{\Delta V}^{\left(n\right)}\\
				\pmb{\beta}_{kn}^{\left(m\right)} = & \left[\left\{w_{0}^{2}\pmb{\mathds{1}} + \pmb{A}^{\left(m\right)}\right\}^{-1}\right]_{kn}\\
				\pmb{A}_{kn}^{\left(m\right)} = & w_{k}w_{n}\left[\pmb{\Delta F}^{\left(n\right)}\right]^{\dagger}\pmb{F}^{\left(k\right)},
			\end{split}
		\end{equation}\par\noindent
		where $\pmb{\mathds{1}}$ is an $\left(m - 1\right)\times\left(m - 1\right)$ identity matrix, $\alpha\in[0, 1]$ is the mixing parameter and $w_{n}$ is the weight associated with the $n$\textsuperscript{th} iteration. We choose $w_{n} = 1, n\neq 0$ and $w_{0} = 0.01$, following previous literature \cite{johnson1988modified, baran2008broyden, vzitko2009convergence}.
	\bibliographystyle{apsrev4-2}
	\bibliography{ref}
\end{document}